\documentclass[journal,article,accept,moreauthors,pdftex,12pt,a4paper]{mdpi} 
\lastpage{x}
\doinum{10.3390/------}
\pubvolume{xx}
\pubyear{2014}
\history{Received: xx / Accepted: xx / Published: xx}

\Title{Measuring the Complexity of Self-organizing Traffic Lights}

\Author{Dar\'io Zubillaga$^{1,2}$, Geovany Cruz$^{1,3}$, Luis Daniel Aguilar$^{1,3}$, Jorge Zapot\'ecatl$^{1,2}$, Nelson Fern\'andez$^{4,5}$, Jos\'e Aguilar$^{5}$, David A. Rosenblueth$^{1,6}$ ,  Carlos Gershenson$^{1,6,}$*}

\address{%
$^{1}$ Departamento de Ciencias de la Computaci\'on,
Instituto de Investigaciones en Matem\'aticas Aplicadas y en Sistemas,
Universidad Nacional Aut\'onoma de M\'exico.\\
$^{2}$ Posgrado en Ciencia e Ingenier\'ia de la Computaci\'on,
Universidad Nacional Aut\'onoma de M\'exico.\\
$^{3}$ Tecnol\'ogico de Estudios Superiores de Jocotitl\'an, M\'exico.\\
$^{4}$ Laboratorio de Hidroinform\'{a}tica, Facultad de Ciencias B\'{a}sicas,
Univesidad de Pamplona, Colombia\\
$^{5}$ Centro de Micro-electr\'onica y Sistemas Distribuidos,
Universidad de los Andes, M\'erida, Venezuela\\
$^{6}$ Centro de Ciencias de la Complejidad,
Universidad Nacional Aut\'onoma de M\'exico.}

\corres{cgg@unam.mx}

\abstract{We apply measures of complexity, emergence and self-organization to an abstract city traffic model for comparing a traditional traffic coordination method with a self-organizing method in two scenarios: cyclic boundaries and non-orientable boundaries. We show that the measures are useful to identify and characterize different dynamical phases. It becomes clear that different operation regimes are required for different traffic demands. Thus, not only traffic is a non-stationary problem, which requires controllers to adapt constantly. Controllers must also change drastically the complexity of their behavior depending on the demand. Based on our measures, we can say that the self-organizing method achieves an adaptability level comparable to a living system.}


\begin{document}


\section{Introduction}

We live in an increasingly urban world~\cite{Cohen:2003,Cities:2010,Roberts:2011}. Cities offer several advantages and thus attract population~\cite{Glaeser:2011,Bettencourt:2007,Bettencourt:2010}. However, this growth also generates problems for which we do not have a clear solution. One of these problems is mobility~\cite{Gyimesi:2011}, which in itself has several factors~\cite[pp. 404--405]{Gershenson:2013}. Efficient traffic light coordination relates to some of these factors, as it can increase the mobility capacity by changes in infrastructure and technology.

Due to its inherent complexity, traffic varies constantly~\cite{Gershenson:2011e}, as vehicles, citizens, and traffic lights interact. As most urban systems, traffic is non-stationary~\cite{Gershenson:2011b}. For this reason, several adaptive approaches to traffic light coordination have been proposed~\cite{FHA2005,Henry:1983,Mauro:1990,Robertson:1991,FaietaHuberman1993,Gartner:2001,Diakaki:2003,FouladvandEtAl2004a,Mirchandani:2005,Bazzan2005,HelbingEtAl2005,Gershenson2005}. Many of these have shown considerable improvements over static methods, which is understandable due to the non-stationary nature of the problem. Some of these proposals can be described as using self-organization to achieve traffic light coordination. A question remains: \emph{how should self-organization be guided to achieve efficient traffic flow?} This can be explored under the nascent field of guided self-organization~\cite{Prokopenko:2009,Ay2012Guided-self-org,GSO2013,GSOInception2014}.

Non-stationary systems change constantly. Which should be their desired regime? Would this also change? How can these be measured? We explore these questions in the context of traffic light coordination, applying recently proposed measures of complexity, emergence and self-organization based on information theory~\cite{Fernandez2013Information-Mea}. In the next section we present our working framework: a city traffic model, the traffic light coordination methods compared, and the proposed measures. Results follow under different boundary conditions. Discussion, future work, and conclusions close the paper.


\section{Methods}

We performed our study on a simulation developed in NetLogo~\cite{Wilensky1999}, available at \url{http://tinyurl.com/trafficCA} including its source code. In the next subsections, we present our traffic model, the traffic light coordination methods compared, and the novel measures used on them.

\subsection{Traffic Model}

We used a previously proposed city traffic model~\cite{RosenbluethGershenson:2010} based on elementary cellular automata~\cite{WuenscheLesser1992,Wolfram:2002}. This is an abstract model (deterministic, time and space are discrete, velocity is either one or zero, acceleration is infinite) which allows us to identify clearly different dynamical phases. The purpose of this model is not predictive, but descriptive~\cite{RosenbluethGershenson:2010}.

\begin{figure}[htbp]
\begin{center}
\includegraphics[width=0.75 \linewidth]{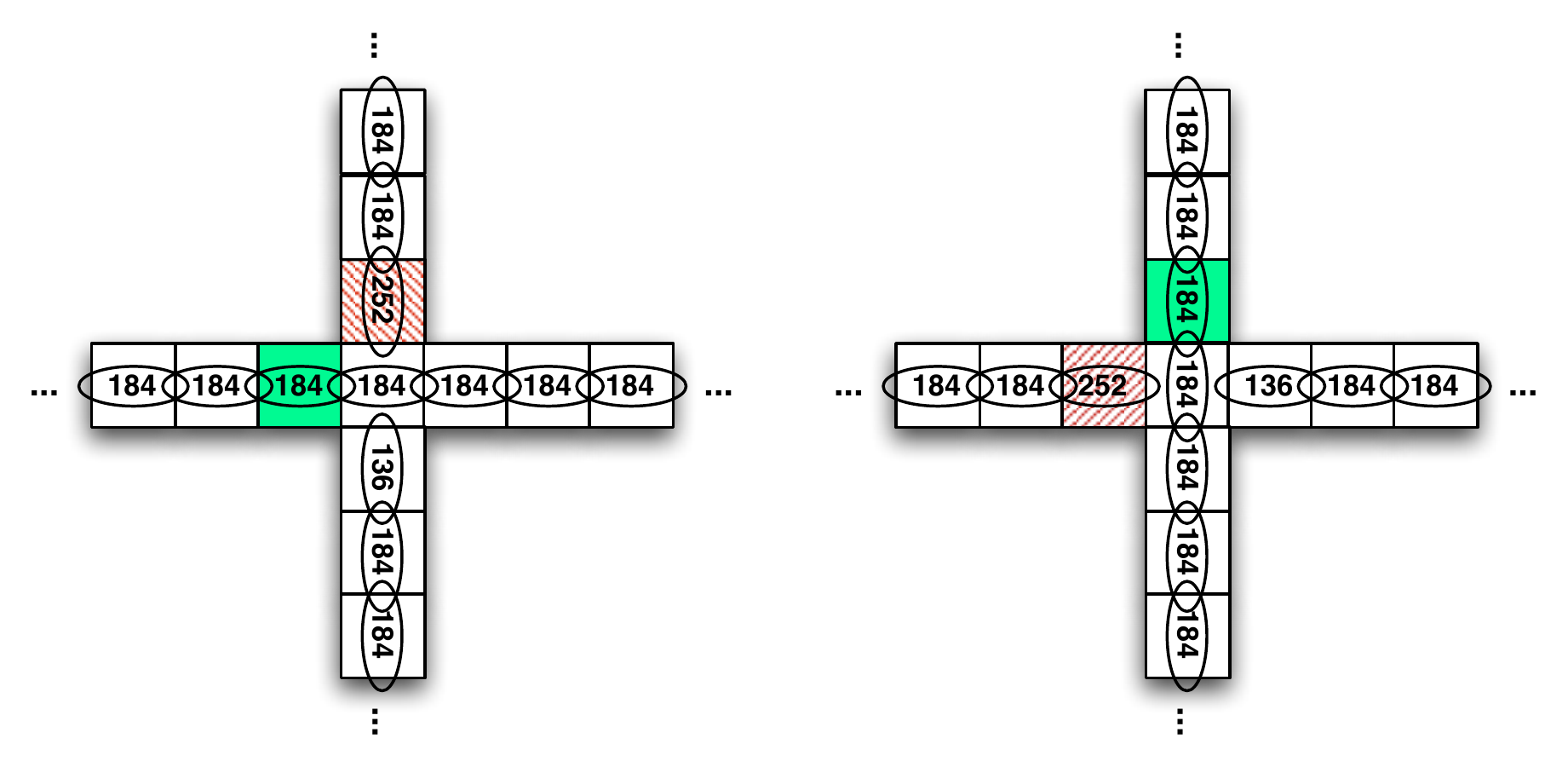}
\caption {ECA rules used in city traffic model as traffic lights switch~\cite{RosenbluethGershenson:2010}.}
\label{fig:rulesDiagram}
\end{center}
\end{figure}

Each street is represented as an elementary cellular automaton (ECA), coupled at intersections. Each ECA contains a number of cells which can take values of zero (empty) or one (vehicle). The state of cells is updated synchronously taking into consideration their previous state and the previous state of their closest neighbors. Most cells simply allow a vehicle to advance if there is a free space ahead. This behavior is modeled with ECA rule 184 (See Table~\ref{table:ECArules}). At intersections, two other rules are used. Before a red light, vehicles do not advance, achieved with ECA rule 252. After a red light, crossing vehicles should not enter the street, achieved with ECA rule 136. The street with the green light uses only ECA rule 184. The intersection cells change neighbourhood to cells on the street with the green light, as shown in Figure~\ref{fig:rulesDiagram}. 

\begin{table}[htdp]
\caption{ECA rules used in city traffic model~\cite{RosenbluethGershenson:2010}. All possible state values of three cells (self and closest neighbors) at time $t-1$ determine the value of the cell at time $t_{rule}$.}
\begin{center}
\begin{tabular}{|c|c|c|c|}
\hline
$t-1$	&$t_{184}$	&$t_{252}$	&$t_{136}$\\
\hline
000	&0	&0	&0	\\
\hline
001	&0	&0	&0	\\
\hline
010	&0	&1	&0	\\
\hline
011	&1	&1	&1	\\
\hline
100	&1	&1	&0	\\
\hline
101	&1	&1	&0	\\
\hline
110	&0	&1	&0	\\
\hline
111	&1	&1	&1	\\
\hline
\end{tabular}
\end{center}
\label{table:ECArules}
\end{table}%

This city traffic model is conservative, i.e. the density of vehicles $\rho$ (percentage of cells with a value of one) is constant in time. It is straightforward to extend it to an hexagonal grid, allowing for more complex intersections~\cite{GershensonRosenblueth:2011}.

The average velocity $v$ is straightforward to calculate: it is given by the number of cells that changed from zero to one (movement occurred between time $t$ and $t+1$) divided by the total number of vehicles (cells with one). $v=0$ if no vehicle moves, while $v=1$ when all vehicles are moving.

Flow $J$ is defined as the velocity $v$ multiplied by the density $\rho$, i.e. $J=v\cdot \rho$. $J=0$ when there is no flow: either there are no vehicles in the simulation ($\rho=0$) or all vehicles are stopped ($v=0$). A maximum $J$ occurs  when all intersections are being crossed by vehicles. In the studied scenarios, $J_{max}=0.25$.

Theoretically, the optimum $v$ and $J$ for a given $\rho$ in a traffic light coordination problem should be the same as optimal $v$ and $J$ for isolated intersections, i.e. an upper limit. This implies that each intersection interacting with its neighbors is as efficient as one without interactions. This can be shown with ``optimality curves"~\cite{GershensonRosenblueth:2011}, visually and analytically comparing results of different methods with theoretical optima.

\subsubsection{Non-orientable boundaries}

Our previous studies have assumed cyclic boundaries, as it is usual with ECA. However, this can lead to a quick stabilization of the system for certain densities, as vehicle trajectories become repetitive (assuming no turning probabilities, the model is deterministic).

Following a suggestion by Masahiro Kanai~\cite{Kanai:2010}, we changed the boundary conditions to be non-orientable (as in a M\"obius strip or a Klein bottle), maintaining incoming vehicle distributions, constant density, and determinism but in practice destroying the correlations that were formed with cyclic boundaries. 

\begin{figure}[htbp]
\begin{center}
\includegraphics[width=0.95 \linewidth]{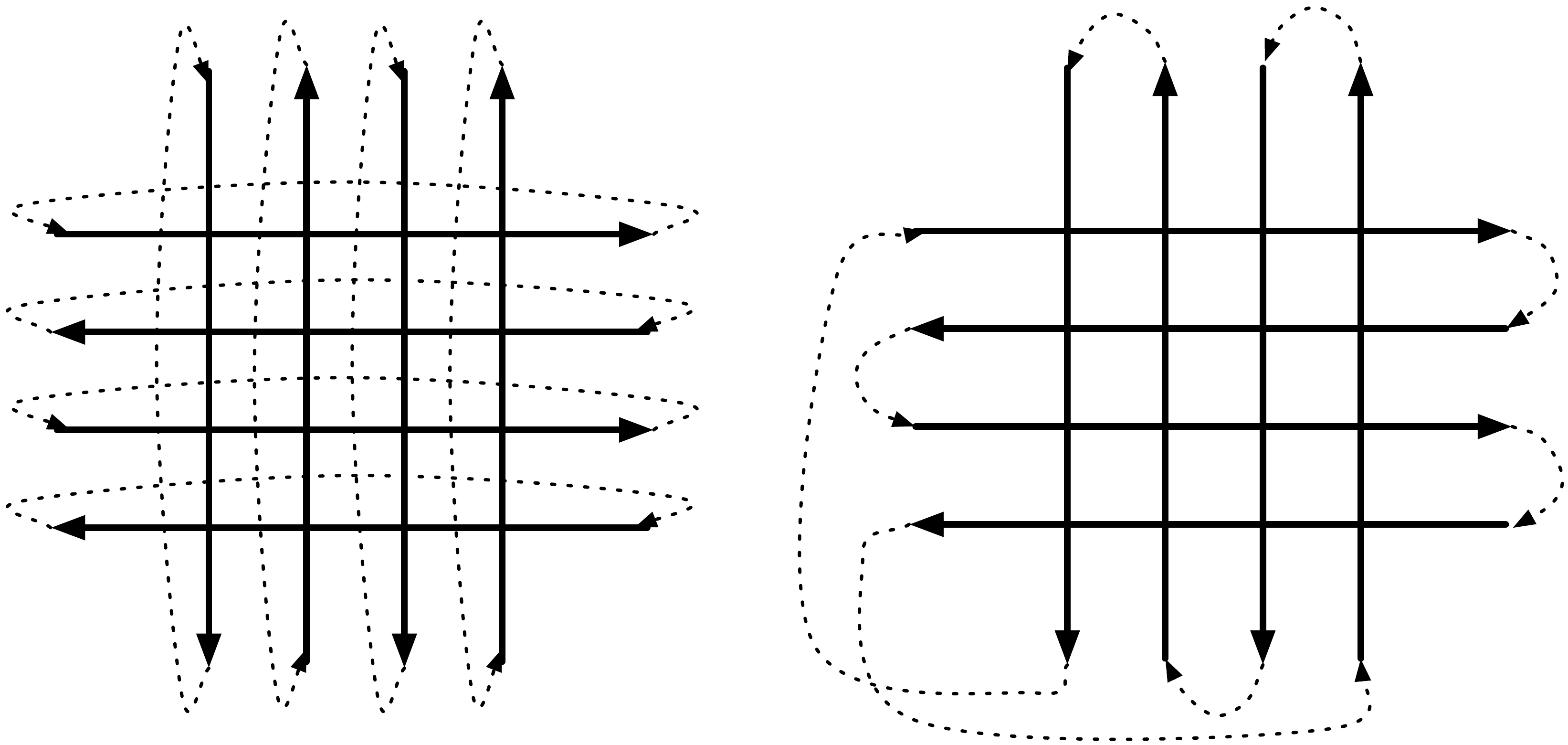}
\caption {Cyclic and non-orientable boundaries.}
\label{fig:boundaries}
\end{center}
\end{figure}

Kanai~\cite{Kanai:2010} studied an isolated intersection, also using ECA. Instead of having two cyclic ECA, he used a single self-intersecting ECA where vehicles exiting on the east enter from the north and vehicles exiting on the south enter from the west. Using a similar approach, changing the boundaries we transformed a ten by ten street scenario (twenty cyclic streets with one hundred intersections) to a single street self-intersecting a hundred times. An example of this change of boundaries is shown in Figure~\ref{fig:boundaries}.

\subsection{Traffic Light Coordination}

The problem of coordinating traffic lights is EXPTIME-complete~\cite{PapadimitriouTsitsiklis1999}. This implies that optimization for large traffic networks is unfeasible. Moreover, the precise traffic conditions change constantly (each cycle a different number of vehicles arrive at each intersection), so the problem is \emph{non-stationary}. Even if we find an optimal solution, this becomes obsolete in seconds. Thus, an alternative to optimization becomes adaptation, being self-organization a useful method for building adaptive systems~\cite{GershensonDCSOS}.

In this work we compare a traditional method which tries to optimize expected flows and a self-organizing method. These are described in detail in~\cite{GershensonRosenblueth:2010} and have been replicated in a parallel implementation, also available with source code at \url{https://github.com/Zapotecatl/Traffic-Light}

\subsubsection{Green-wave method}

Most cities try to synchronize their traffic lights using the so-called green-wave method~\cite{TorokKertesz1999}. Traffic lights have the same period and the phase (offset) is adjusted so that green lights switch at the expected velocity of vehicles. In this way, once vehicles get a green light, they should not get any red light. This is better than having no coordination. However, due to mathematical constrains, at most two directions can have a green wave at the same time. Thus, vehicles driving in the opposite direction find anti-correlated phases and have high waiting times, leading to long queue formation already for medium densities. Moreover, if the vehicles do not go at the expected velocity (as it is the case when densities change), then vehicles will not be able to go at the speed of the green wave and will have to stop.

In~\cite{GershensonRosenblueth:2010} we studied the green wave method in the traffic model described in the previous subsection. We found only two dynamical phases: \emph{intermittent} (some vehicles stop, some move) and \emph{gridlock} (all vehicles are stopped).

\subsubsection{Self-organizing method}

We have proposed and refined a self-organizing method for traffic light coordination where each intersection follows simple local rules, reaching close to optimal global coordination~\cite{Gershenson2005,CoolsEtAl2007,GershensonRosenblueth:2010,GershensonRosenblueth:2011}. The algorithm follows six simple rules, listed in Table~\ref{table:rules}. Rules with higher numbers override rules with lower numbers, e.g. rule 4 overrides rule 1.

\begin{table*}[htdp]
\caption{Six rules of the self-organizing method~\cite{GershensonRosenblueth:2010}. Inset: Schematic of an intersection, indicating distances $d$, $r$, and $e$ used.}
\label{table:rules}
\begin{center}
\vspace{0.2cm}
\fbox{
\parbox{.9\textwidth}{

 \begin{center}
   \includegraphics[height=30mm]{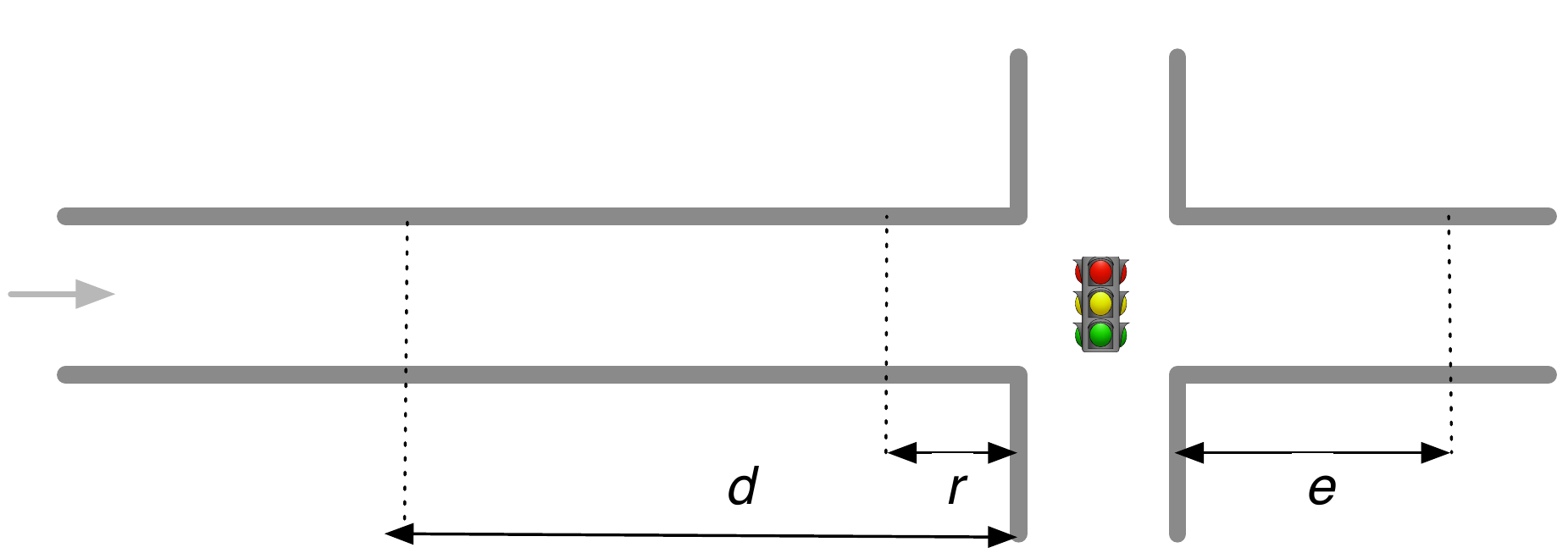}
 \end{center}

{\small
\begin{enumerate}
\item On every tick, add to a counter the number of vehicles
approaching or waiting at a red light within distance $d$.
When this counter exceeds a threshold $n$, switch the light.
Whenever the light switches, reset the counter to zero.
\item Lights must remain green for a minimum time $u$.
\item If a few vehicles ($m$ or fewer, but more than zero) are left to cross a green light at a short distance $r$, do not switch the light.
\item If no vehicle is approaching a green light within a distance $d$, and at least one vehicle is approaching
the red light within a distance $d$, then switch the light.
\item If there is a vehicle stopped in the street a short distance $e$
  beyond a green traffic light, then switch the light.
\item If there are vehicles stopped on both directions at a short distance $e$
  beyond the intersection, then switch both lights to red. Once one of the directions is free, restore the green light in that direction.
\end{enumerate}

}

}

}

\end{center}

\end{table*}%

Rule 1 gives preference to streets with higher demand (few vehicles wait more than several) and promotes the formation of platoons (few vehicles waiting may be joined by more to form larger groups, which reaching further intersections can trigger the green light before decreasing their speed). Rule 2 maintains a minimum green time to prevent fast switching of traffic lights in high densities. Rule 3 maintains platoons together, although allowing to split large platoons. Rule 4 is useful for low densities, allowing few vehicles to trigger a green light if there is no vehicle approaching the current green. Rules 5 and 6 are useful for high densities, switching lights to red if there are cars stopped downstream of the intersection, preventing its blockage. The pseduocode of the algorithm extended for multiple directions can be found in~\cite{GershensonRosenblueth:2011}.

In~\cite{GershensonRosenblueth:2010} we found seven dynamical phases of the self-organizing method in our traffic model (with cyclic boundaries) as the density $\rho$ increases: \emph{free flow} (no vehicle stops, $v=1$), \emph{quasi-free flow} (almost no vehicle stops), \emph{intermittent underutilized} (some vehicles stop, intersections do not reach maximum flow $J_{max}$), \emph{full capacity intermittent} (some vehicles stop, $J_{max}$ i.e. all intersections have vehicles using them always), \emph{overutilized intermittent} (some vehicles stop, intersections have to restrict both streams to prevent blockages using rule 6), \emph{quasi-gridlock} (almost all vehicles are stopped, but ``platoons" of spaces form and move in the direction opposite of streets), and \emph{gridlock} (all vehicles are stopped, i.e. $v=0$, as intersections get blocked from initial conditions). Their respective phase transitions occur approximately at $\rho$ values of $0.15, 0.22, 0.38, 0.63, 0.77,$ and $0.95$.

\subsection{Measures}

We recently proposed measures of emergence, self-organization, and complexity based on information theory in~\cite{GershensonFernandez:2012} and refined them and based them on axioms in~\cite{Fernandez2013Information-Mea}.

Shannon defined a measure of information~\cite{Shannon1948} equivalent to Boltzmann's entropy depending of the probabilities $p_i$ for all $i$ symbols in a finite alphabet:
\begin{equation} 
I = -K \sum_{i=i}^{n} p_{i} \log p_{i}, 
\label{eq:I}
\end{equation}
where $K$ is a positive constant. In this work, we consider the $\log$ in eq.~\ref{eq:I} to be of base two.

In its most general form, emergence can be understood as information produced by a process or system. Shannon's information already measures this, so we defined
\begin{equation}
E=I.
\label{eq:E}
\end{equation}

Minimum $E=0$ is given for regular, predictable strings (no new information produced), while maximum $E=1$ is given for irregular, pseudorandom strings (each symbol is new information). To bound $E$ to the interval $[0,1]$ we simply consider
\begin{equation}
K = \frac{1}{\log_{2}b},
\end{equation}
where $b$ is the base used, i.e. the length of the alphabet (possible symbols which can occur in a string). For example, if $b=2$, then equation~\ref{eq:I} can be rewritten as:
\begin{equation} 
I = -(p_{0} \log_2 p_{0} + p_{1} \log_{2} p_{1}),
\label{eq:I}
\end{equation}
since $K=1$. In this work, we use base ten, i.e. $b=10$. In our experience, a slight change of $b$ does not affect qualitatively the measures.

Self-organization can be seen as a reduction of entropy~\citep{GershensonHeylighen2003a}. Since $E$ is a type of entropy, we define 
\begin{equation}
S = 1 - E = 1 - I.
\label{eq:S}
\end{equation}

It might seem counterintuitive to define self-organization as the opposite of emergence, being both properties of complex systems. However, when taken to their limits, it can be seen that emergence is maximal in chaotic systems ($E=1$, $S=0$) and self-organization is maximal in ordered systems ($S=1$, $E=0$). It is when these are balanced that we can have complexity. Following L\'opez-Ruiz et al.~\cite{LopezRuiz:1995}, we define
\begin{equation} 
C = 4 \cdot E \cdot S = 4 \cdot I \cdot (1-I),
\label{eq:C}
\end{equation}
where the 4 is included as a normalizing constant, bounding $C$ to $[0,1]$. 

Historically, Shannon defined information as entropy~\cite{Shannon1948} (equivalent to our $E$), while Wiener~\cite{Wiener1948} and von Bertanalffy~\cite{von-Bertalanffy:1968} defined information as its oposite, negentropy (equivalent to our $S$). Our measure of complexity $C$ reconciles these two opposing views, as a balance between order and chaos is maximal with a high $C$. Dynamical systems such as cellular automata~\cite{Langton1990} and random Boolean networks~\cite{Kauffman1969,Kauffman1993,Gershenson2004c} have a maximal $C$ in the region their dynamics are considered most complex~\cite{GershensonFernandez:2012}. Since living systems also require a balance between adaptivity ($E$) and stability ($S$)~\cite{Kauffman1993,Balleza:2008}, $C$ can be used to characterize living systems, especially when comparing their $C$ with that of their environment~\cite{Fernandez2013Information-Mea}. If we want artificial systems to exhibit the properties of living systems~\cite{Bedau:2009,Bedau2013IntroductionLT,Gershenson:2013}, then these should also have a high $C$ compared with that of their environment.

In this work, we measure the $E$, $S$, and $C$ of three different time series: of switching intervals, of vehicle intervals at an intersection, and of vehicle intervals at a street. These series measure the time between light switches of vehicles crossing a point. If these are completely regular, then $E=C=0$ and $S=1$. If these were maximally chaotic, then $S=C=0$ and $E=1$.


\section{Results}

Results in this section were obtained simulating a 10x10 street grid, each equidistant street representing 800m, i.e. 80m per block. A cell represents 5m (the distance between the front ends of stopped vehicles), and a time step one third of a second. Thus, $v=1$ represents 54km/h.

Simulations were performed for one hundred densities $\rho$ between 0.01 and 1, fifty runs per density value (5000 runs total with random initial conditions). Standard deviations are shown in figures, although these were minimal, except for some regions near phase transitions. Each run consisted of ten thousand time steps, of which only the second half were considered for generating the statistics.

All figures in the Results and Discussion sections show phase transitions of the self-organizing method for the cyclic boundary scenario with dotted lines for comparison.

\subsection{Cyclic boundaries}

\begin{figure}
     \centering
      \includegraphics[width=.9\textwidth]{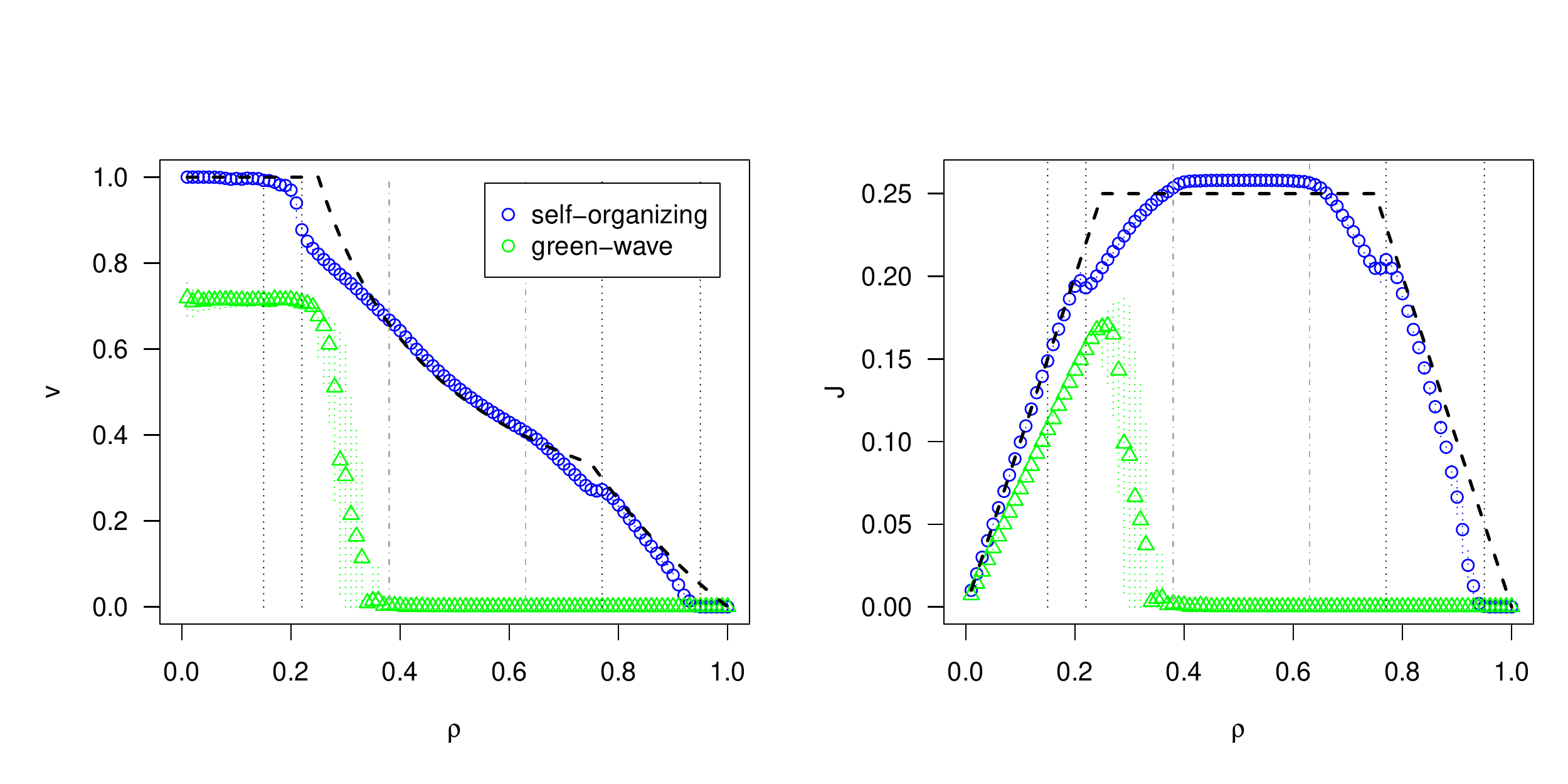}    
     \caption{Results for cyclic boundaries: average velocity $\langle v\rangle$ and average flux $\langle J\rangle$ for different densities $\rho$. Optimality curves shown with dashed black lines. Phase transitions of the self-organizing method are indicated with vertical dotted lines.}
     \label{fig:cyclic:vJ}
\end{figure}

As reported in~\cite{GershensonRosenblueth:2010}, the performance---based on $v$ and $J$---of the self-organizing method is superior to the green-wave method for all densities, as it can be seen in Figure~\ref{fig:cyclic:vJ}. Here we include optimality curves~\cite{GershensonRosenblueth:2011}, showing that the self-organizing method achieves or reaches a performance close to the theoretical optimum for a given density. The full capacity intermittent phase is optimal by definition, as well as the free flow phase. It can be seen that small improvements can still be made in the underutilized and overutilized phases, as well as the quasi-gridlock and gridlock phases.

\begin{figure}
     \centering
      \includegraphics[width=.98\textwidth]{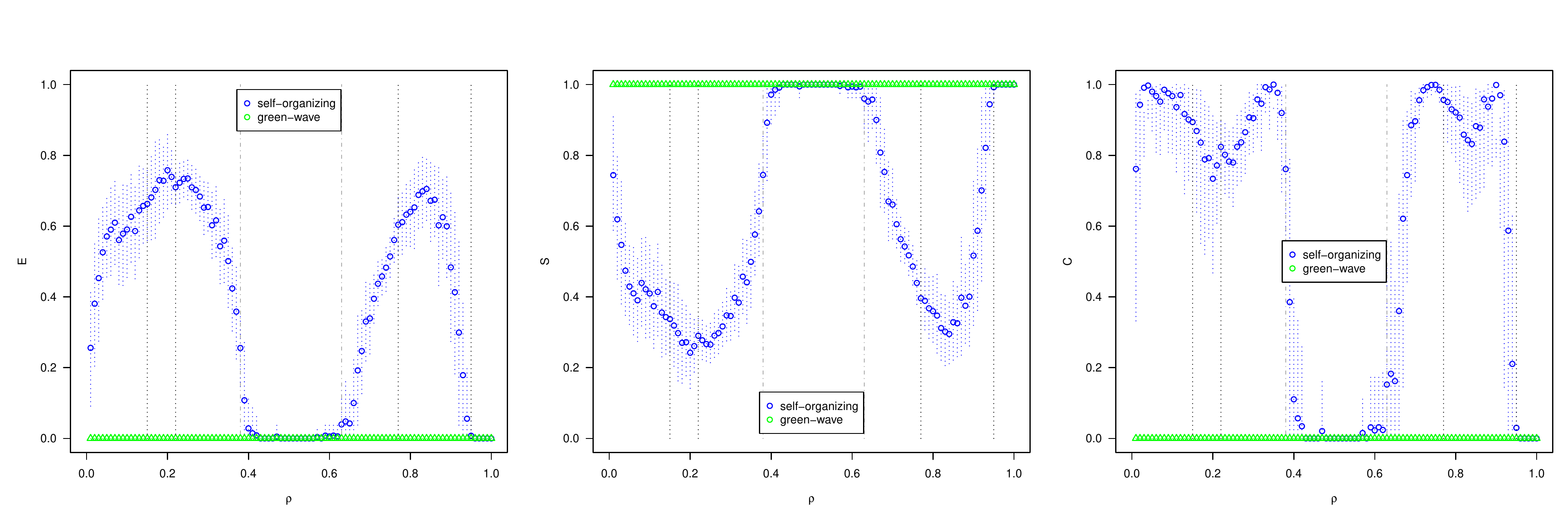}    
     \caption{Results for cyclic boundaries: $E$, $S$, and $C$ of switching intervals for different densities $\rho$.}
     \label{fig:cyclic:swI}
\end{figure}

Figure~\ref{fig:cyclic:swI} shows results of measures for switching intervals of intersections.

Since the green wave method has periodic cycles, for their switching $E=C=0$ and $S=1$. Note that our $S$ measure does not distinguish whether the organization is internal (self-) or external, as it is the case for this method which depends on a central controller. Having extreme $S=1$, it cannot adapt to changes in traffic flow. 

The self-organizing method adapts constantly to changes in demand, as it can be seen from the measures variation for different densities. The switching is most irregular (greatest $E$, lowest $S$) in the quasi-free flow and quasi-gridlock phases, while it achieves a regular switching (minimal $E$ and $C$, maximal $S$) for the full-capacity intermittent phase, as there is always demand to be served from all directions. $C$ is high for all phases but full-capacity intermittent and gridlock, exhibiting similar measures as the green wave method for these two phases.

\begin{figure}
     \centering
      \includegraphics[width=.98\textwidth]{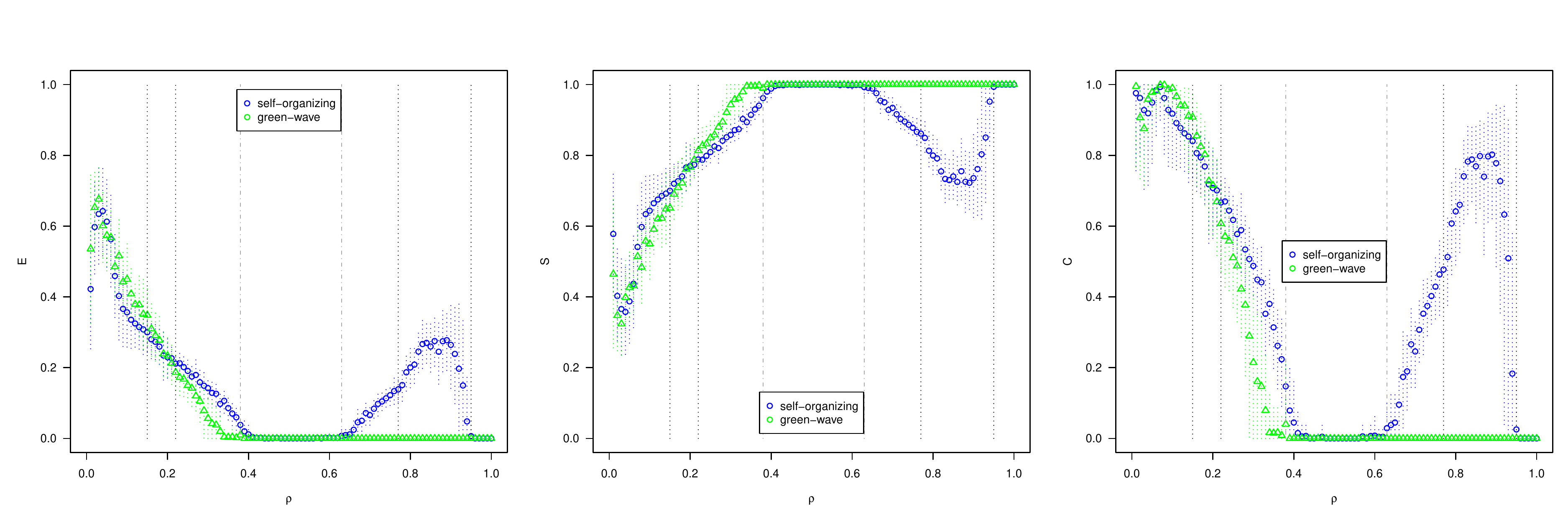}    
     \caption{Results for cyclic boundaries: $E$, $S$, and $C$ of vehicle intervals at an intersection for different densities $\rho$.}
     \label{fig:cyclic:ciI}
\end{figure}

The measures of vehicles crossing at an intersection, as seen in Figure~\ref{fig:cyclic:ciI}, are similar for the full capacity intermittent and gridlock phases, as there are either vehicles constantly crossing the intersection (there is a space between moving vehicles, so the intervals are constantly two time steps) or no vehicles moving (gridlock). $E$ and $C$ are high for low densities for both methods, and these become more regular ($S$ increases) with $\rho$ until reaching a maximum in the full capacity intermittent phase for the self-organizing method and the gridlock phase in the green wave method. The self-organizing method increases again $E$ and $C$ until the quasi-gridlock phase, as vehicle crossings become less regular and ``space platoons" are formed, only to decrease again towards the gridlock phase.

Vehicle intervals are measured at different street locations (not intersections) chosen randomly each run. These measures are similar to those from intersections described above, although they reach maximum $S=1$ only for the gridlock phase. There seems to be a minimum $E\approx0.2$ for the full capacity intermittent phase. Long platoons are formed, so most crossing intervals are two. However, between two platoons there are long spaces which imply a single long interval between the last vehicle of a platoon and the first one of the next platoon, giving minimum irregularity to the time series. Notice that $C$ is high for all phases except gridlock, for both methods, although differences can be easily identified between some phases of the self-organizing method. In other words, the dynamical phase can have a direct impact on $E$, $S$, and $C$, which potentially could also be used to identify dynamical phases in more realistic traffic models, where transitions are not as crisp as in ours and thus could be difficult to identify.

\begin{figure}
     \centering
      \includegraphics[width=.98\textwidth]{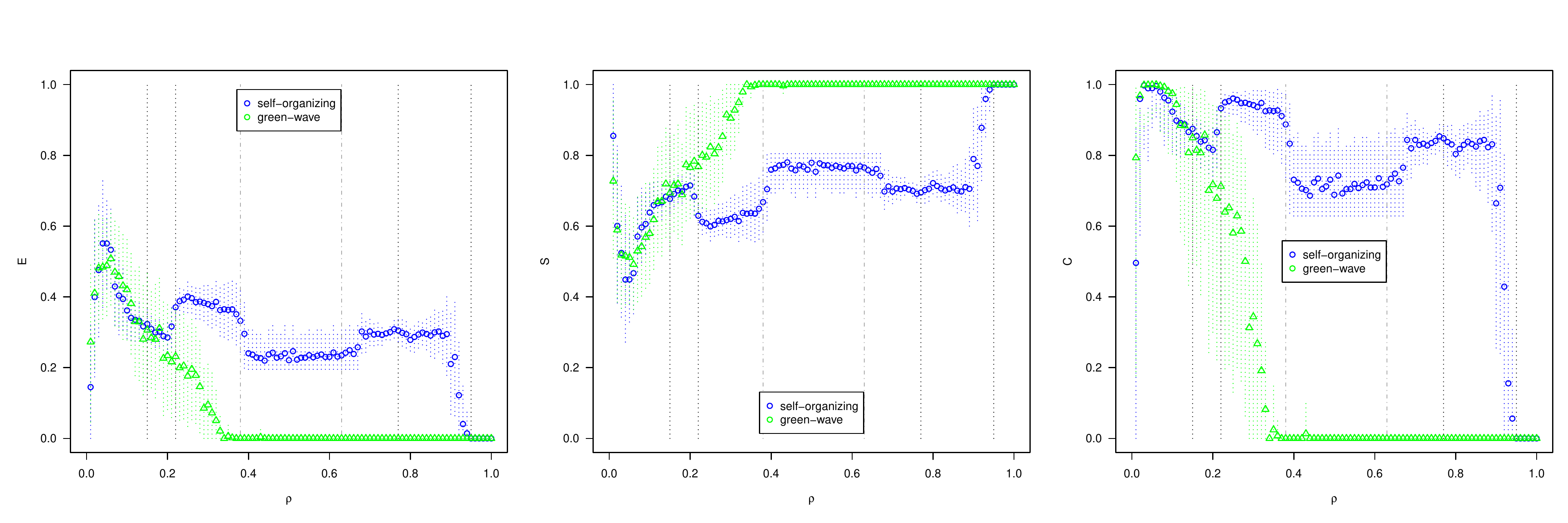}    
     \caption{Results for cyclic boundaries: $E$, $S$, and $C$ of vehicle intervals at a street for different densities $\rho$.}
     \label{fig:cyclic:csI}
\end{figure}

\subsection{Non-orientable boundaries}

\begin{figure}
     \centering
      \includegraphics[width=.9\textwidth]{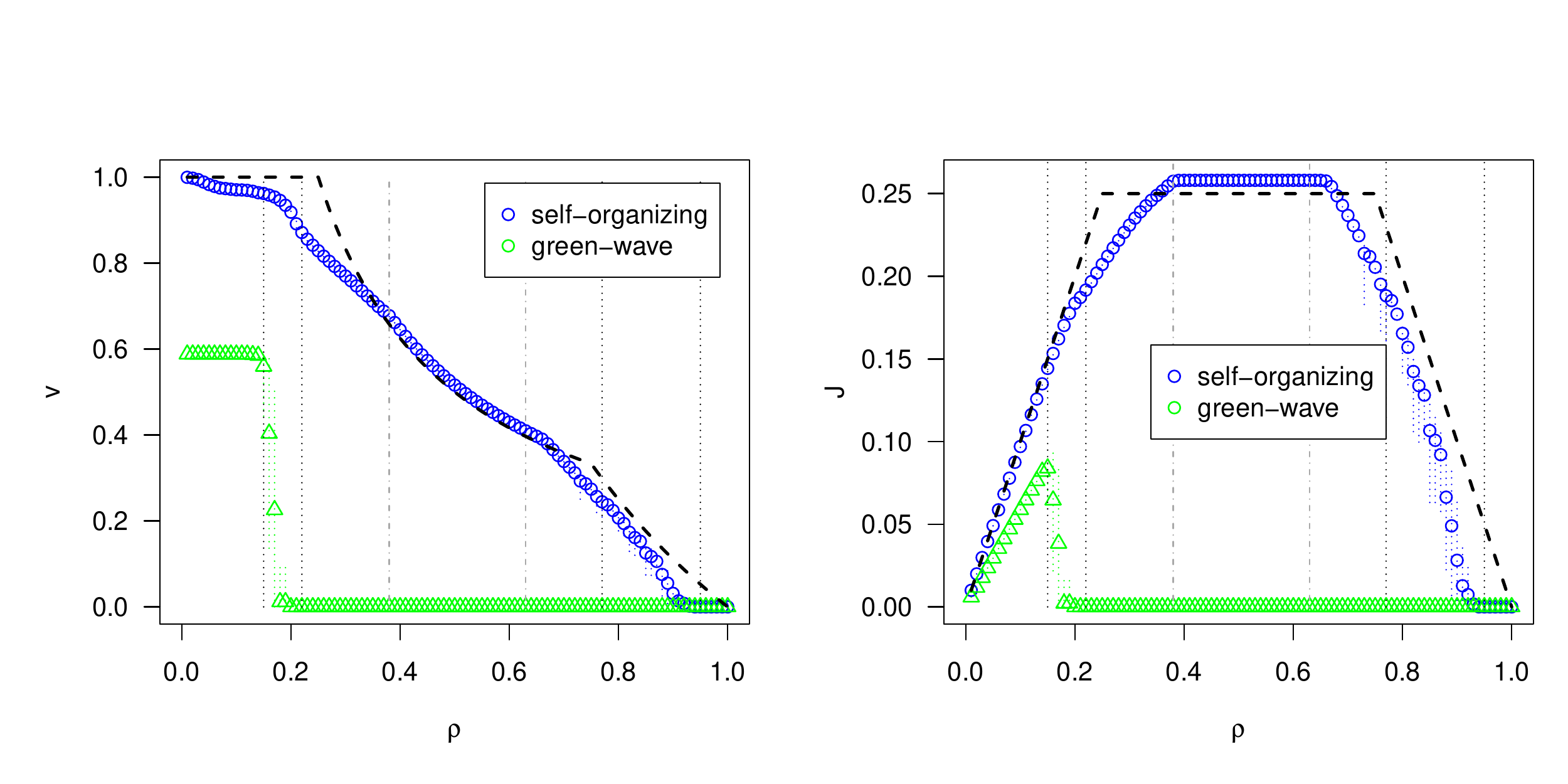}    
     \caption{Results for non-orientable boundaries: average velocity $\langle v\rangle$ and average flux $\langle J\rangle$ for different densities $\rho$. Optimality curves shown with dashed black lines.}
     \label{fig:singleMoebius:vJ}
\end{figure}

We performed similar simulations as the ones described in the previous subsection but with non-orientable boundaries. Results for $v$ and $J$ are shown in Figure~\ref{fig:singleMoebius:vJ}, which indicates with dotted lines the phase transitions of the self-organizing method for the \emph{cyclic} boundaries scenario for comparison. 

Compared to the cyclic boundaries scenario, the green wave method has a lower $v$ for its intermittent phase and transitions into the gridlock phase at a lower density.

For the self-organizing method, there is no free-flow phase, as this depends on the correlations formed by the looping platoons which are destroyed with the non-orientable boundaries. The transition between the quasi-free flow and underutilized intermittent phases occurs at a lower density. The full capacity intermittent phase is expanded for higher densities, transitioning later into the quasi-gridlock phase. The transition between the quasi-gridlock and gridlock phases occurs at a slightly lower density.

In general, the self-organizing method maintains itself at or near the optimality curves, showing that its performance is not strongly dependent on the boundary conditions studied here.

\begin{figure}
     \centering
      \includegraphics[width=.98\textwidth]{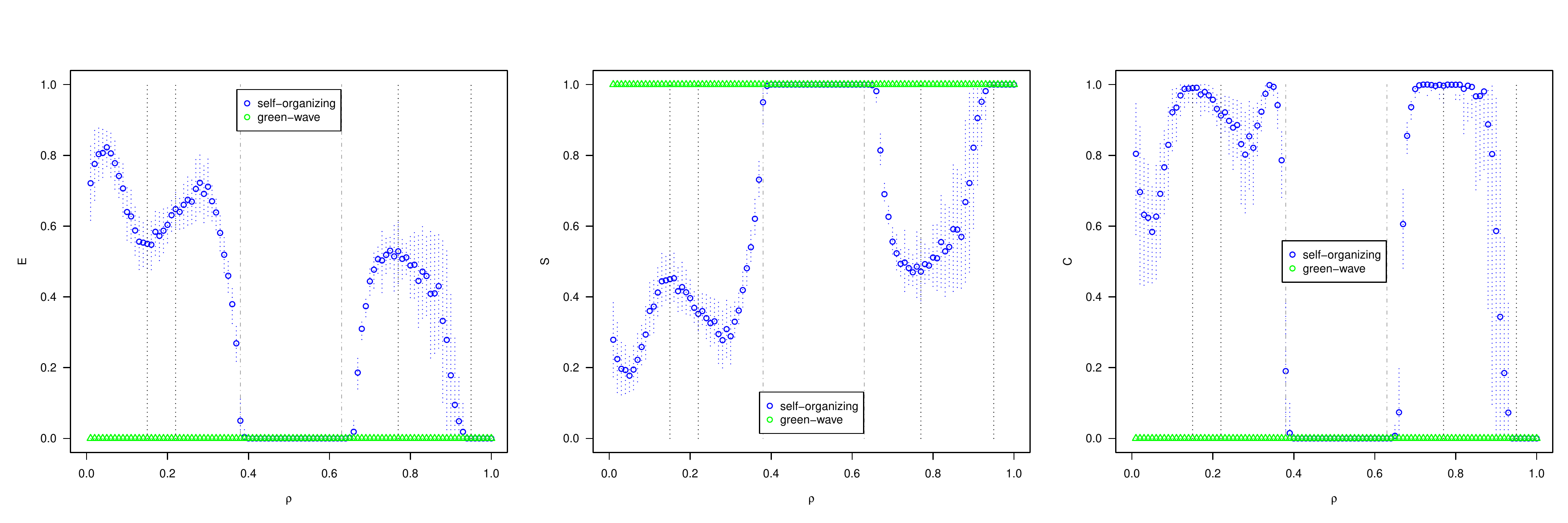}    
     \caption{Results for non-orientable boundaries: $E$, $S$, and $C$ of switching intervals for different densities $\rho$.}
     \label{fig:singleMoebius:swI}
\end{figure}

\begin{figure}
     \centering
      \includegraphics[width=.98\textwidth]{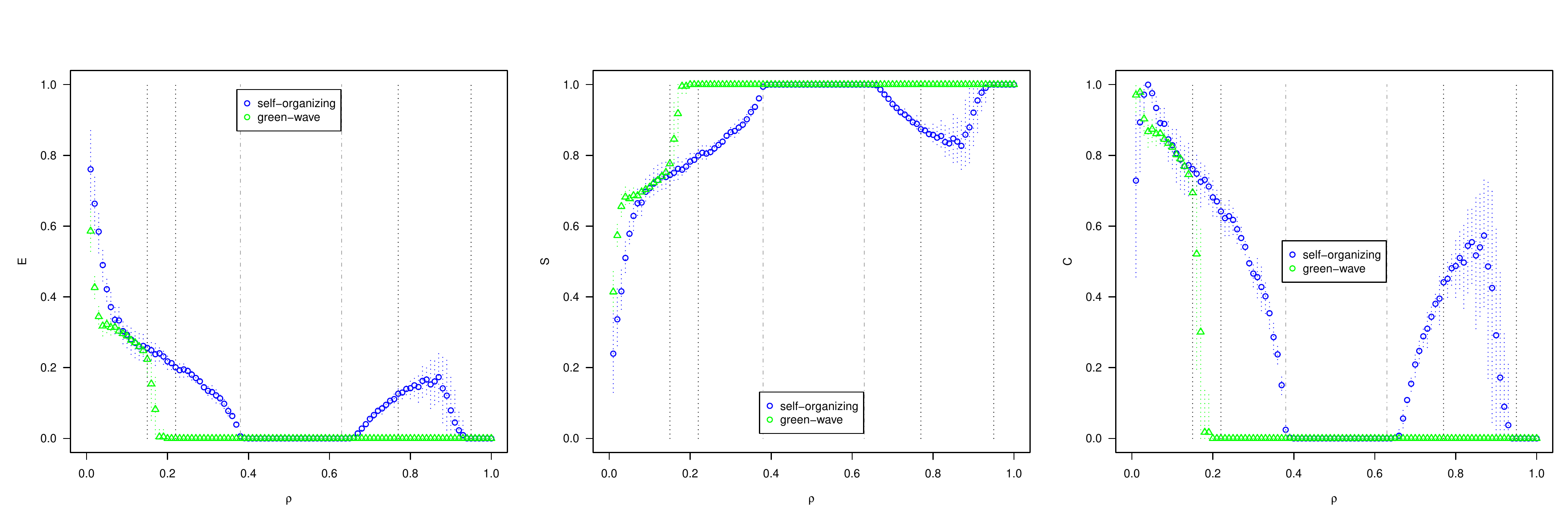}    
     \caption{Results for non-orientable boundaries: $E$, $S$, and $C$ of vehicle intervals at an intersection for different densities $\rho$.}
     \label{fig:singleMoebius:ciI}
\end{figure}

\begin{figure}
     \centering
      \includegraphics[width=.98\textwidth]{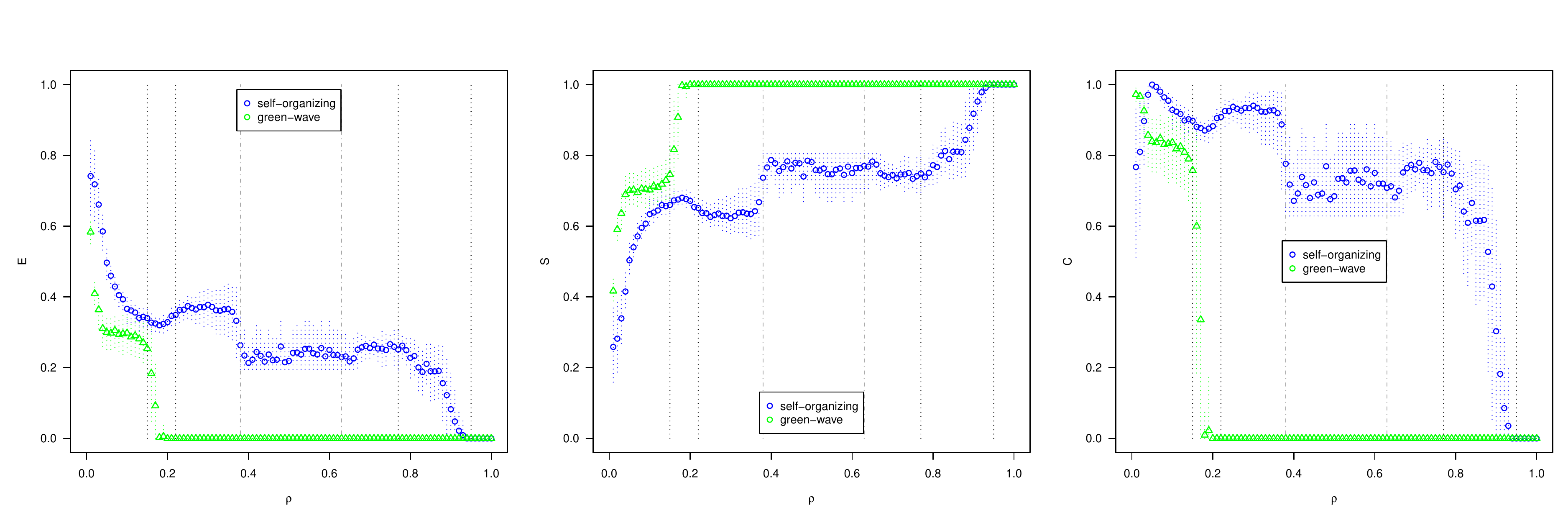}    
     \caption{Results for non-orientable boundaries: $E$, $S$, and $C$ of vehicle intervals at a street for different densities $\rho$.}
     \label{fig:singleMoebius:csI}
\end{figure}

As for the measures of $E$, $S$, and $C$ for switching intervals (Figure~\ref{fig:singleMoebius:swI}), vehicle intervals at intersections (Figure~\ref{fig:singleMoebius:ciI}) and vehicle intervals at streets (Figure~\ref{fig:singleMoebius:csI}), there are no qualitative differences for the same dynamical phases between the cyclic and non-orientable boundary conditions.

\section{Discussion}

Changes in values of $C$, $S$, or $E$ can be used to detect phase transitions, as these can be radically different for different phases. This can also be achieved with other measures, such as Fisher information~\cite{Prokopenko2011Relating-Fisher}. However, it cannot be said whether a specific value of our measures is good or bad independently of density. There are traffic regimes where high regularity ($S$) of traffic lights is required to achieve optimal performance (full capacity intermittent phase), while there are other regimes, where a high adaptability ($C$) is necessary. 
This reflects the non-stationary nature of urban systems~\cite{Gershenson:2013}, where no single solution is efficient for all situations, as these change constantly.

Not only traffic lights have to adapt their duration at the seconds scale to match the scale at which traffic demand changes. Traffic lights have also to adapt their regime at the scale at which density shifts from phase to phase. Extending Ashby's law of requisite variety~\cite{Ashby1956}, it can be said that there is a requisite complexity~\cite{Gershenson:2007} that a controller must have in order to cope with the complexity of the system it attempts to control~\cite[p. 47]{Fernandez2013Information-Mea}. This is similar to the ``matching complexity" proposed by~\cite{Tononi1996}.
Our self-organizing method seamlessly achieves this, giving insights on the requirements of other self-organizing systems which might be used for addressing non-stationary problems.

It might be possible to define measures which capture this requisite complexity and how it should match the complexity of the controlled/environment at different scales. One possibility can be with a measure of \emph{autopoiesis} $A$~\cite{Fernandez2013Information-Mea} which is defined as the ratio between the $C$ of a system (controller) and the $C$ of its environment (controlled). Further work is required in this direction, especially because in the self-organizing method traffic lights control vehicles, but vehicles also control traffic lights to a certain degree.

\begin{figure}
     \centering
      \includegraphics[width=.5\textwidth]{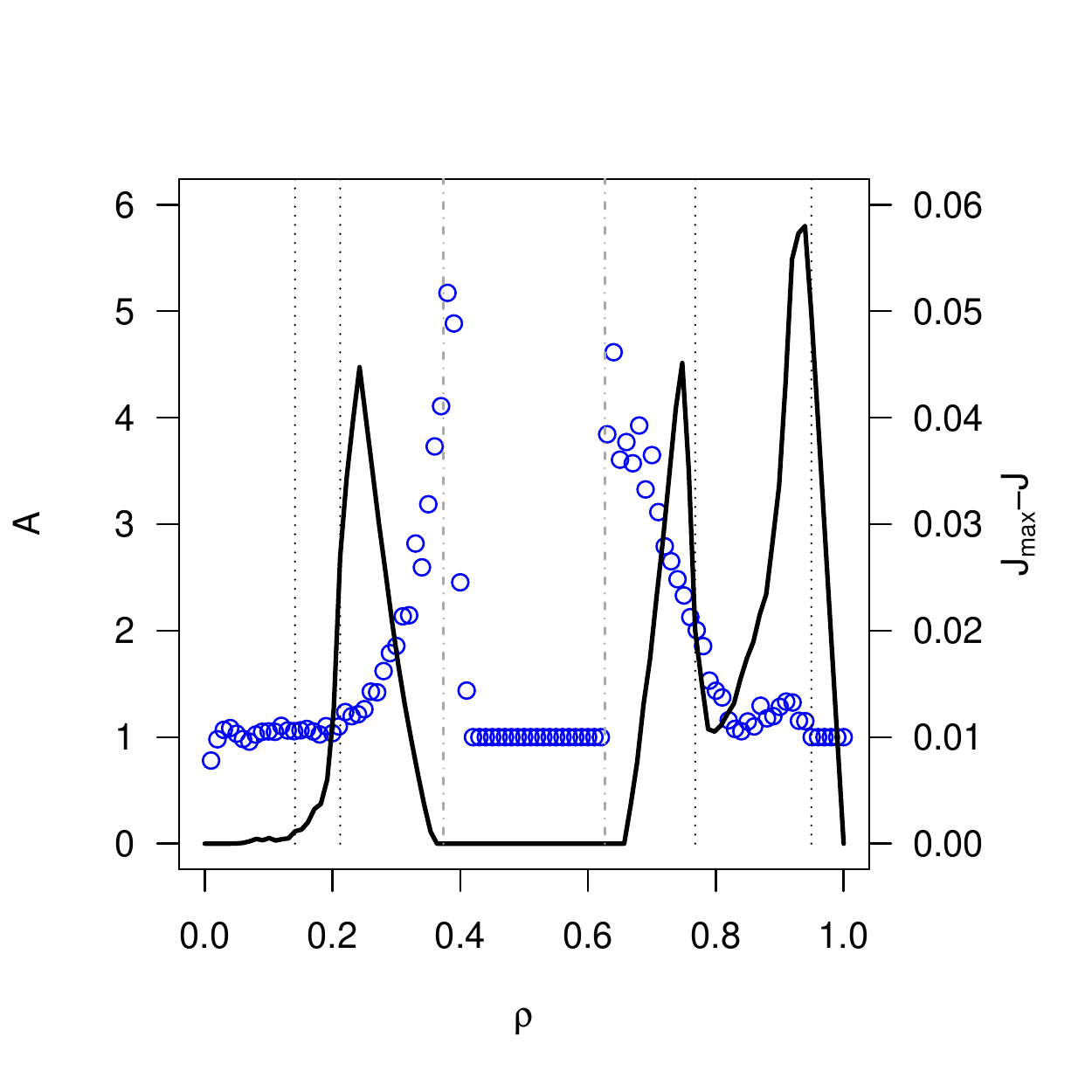}    
     \caption{$A$ of self-organizing method (blue circles) for different densities $\rho$, compared with the $J$ optimality difference (black solid line) for the cyclic boundaries scenario.}
     \label{fig:cyclic:A}
\end{figure}

As an initial exploration, we compared the $C$ ratio between the switching intervals and vehicle intervals at intersections to obtain $A$ for the self-organizing method, shown in Figure~\ref{fig:cyclic:A}.  To avoid the indeterminacy of divisions by zero, only here we imposed a lower bound $C_{min}=0.01$. The switching $C$ is almost always higher or equal than the vehicle $C$ at intersections (not always at streets), leading to $A\geq1$. The highest $A$ values occur close to the phase transitions bordering the full capacity intermittent phase. These regions require a complex switching, and the self-organizing method manages to achieve this to regulate vehicle flow. Figure~\ref{fig:cyclic:A} also includes the difference between the optimality flow curve $J_{max}$ and the actual $J$ of the self-organizing method, to show where the method can still be improved. As mentioned previously, this is near the phase transitions between the quasi-free flow and underutilized intermittent phases and those around the quasi-gridlock phase. It could be speculated that to improve these regions $A$ should be further increased, either by augmenting the $C$ of the switching intervals or by decreasing the $C$ of the vehicle intervals. However, having already $A\geq1$ suggests that the lack of optimality for the specific densities close to the three mentioned phase transitions is due to inherent constrains in the problem. We conjecture that to reach $J_{max}$ around $\rho=0.25$ and $\rho=0.75$ it is necessary to have a homogeneous demand on all streets, so as to prevent ``idling" of some intersections due to some streets having densities different from these values, which do vary in our simulations as initial conditions are random. A similar reasoning can be made for the region around the quasi-gridlock-gridlock phase transition: In order to allow vehicles on a single street with very high $\rho>0.95$ to have $J>0$, initial conditions must be such that all of the intersecting streets must have a space available for the intersection and not others. This is highly unlikely for large grids with random initial conditions.
Both constrains are dependent on the problem and beyond the traffic light coordination method.

From this study, we can generalize the use of our measures for guiding self-organization of other systems. We have seen that a desired value of $C$ depends on the regime which a system requires. If the system benefits from high regularity, then a high $S$ should be sought. If randomness or chaos are desired, then a high $E$ can be covetable. If adaptation is needed, then a high $C$ is advisable. In a control system, from our initial explorations of $A$, a minimum value of the $C$ of the controller will depend on the $C$ of the controlled.


\section{Future Work}

The results presented in this paper are promising, but there are several points that should be further explored. 
\begin{itemize}
	\item The relationship between dynamical phases and measures should be studied in greater detail.
	\item The phase transitions in the non-orientable scenario can be further investigated, relating the performance measures such as $v$ and $J$ with the measures $C$, $E$, and $S$.
	\item We plan to perform similar studies with a more realistic traffic model~\cite{Larraga2010Cellular-automa}, adapted to city traffic.
	\item Massive simulations with hundreds of streets and thousands of intersections are intended to better observe the dynamical phases and their transitions.
	\item We have calculated the optimality curves for $v$ and $J$. What would be the optimality curves for $C$, $S$ and $E$? These cannot be simply extrapolated from the $C$, $S$ and $E$ obtained for an optimal switching of an isolated intersection.
	\item Can our measures be used to guide the self-organization of traffic lights even closer towards optimality?
	\item Can our measures be used to guide the self-organization of other systems?
\end{itemize}


\section{Conclusions}

In this work we applied measures of emergence, self-organization, and complexity to an abstract city traffic model and compared two traffic light coordination methods. Varying boundary conditions yielded similar results. The measures reflect why the self-organizing method is much better than the green-wave method: for certain dynamical phases regular behavior is required, while for others adaptive, complex behavior is most efficient. The green-wave method by definition has only regular behavior, and thus is unable to adapt to constantly shifting demands. The self-organizing method has enough ``requisite complexity" to cope with different behaviors required by changes in demand. Having an autopoiesis greater than one, it can be said that the self-organizing traffic lights have (computationally) the properties of a living system~\cite{Gershenson:2007,Fernandez2013Information-Mea}, which also require to have a greater $C$ than their environment.
Still, there are some regions which are not optimal. Whether these can be reached with a different method or they are computationally unfeasible (and our method then would be optimal in practice) still has to be explored.

	This work shows an example of the usefulness of our proposed measures to guide the self-organization of systems, which are being applied in  other areas as well~\cite{Amoretti:2012,Fernandez_Gershenson_2014,Febres2013Complexity-meas}. 
Since everything can be described as information~\cite{Gershenson:2007}, there is the potential of measuring $C$, $E$, and $S$ of every phenomena. Certainly, interpreting the meaning of a specific value at a particular scale is the real challenge.


\acknowledgements{Acknowledgements}

C.G. was partially supported by SNI membership 47907 of CONACyT, Mexico. 


\conflictofinterests{Conflicts of Interest}

The authors declare no conflicts of interest.

\bibliography{carlos,complex,sos,RBN,traffic,information}
\bibliographystyle{mdpi}

\end{document}